\documentclass{moriond}
\usepackage{amsmath,graphicx,color,epsfig,float}
\usepackage{epstopdf}

\def\be{\begin{equation}}
\def\ee{\end{equation}}
\def\bea{\begin{eqnarray}}
\def\eea{\end{eqnarray}}

\newcommand{\Photo}{%\includegraphics[height=35mm]{photo.png}
}

\begin{document}
\vspace*{4cm}
\title{SEARCH FOR DOUBLY HEAVY BARYON VIA WEAK DECAYS}

\author{Run-Hui Li$^a$, Cai-Dian Lu$^b$ }

\address{$^a$ School of Physical Science and Technology, Inner Mongolia University, Hohhot 010021, China;\\ $^b$Institute of High Energy Physics, YuQuanLu 19B, Beijing 100049, China .}

\maketitle\abstracts{
  Using the factorization approach and taking into account the final state interaction, we calculate the two body non-leptonic decays  of  doubly heavy baryons. After comparing the semi-leptonic decays and all possible hadronic decay channels, we found some channels with large branching ratios. Taking the detection efficiency into consideration, we suggest $\Xi_{cc}^{++}$ as the first search goal and $\Xi_{cc}^{++}\to \Lambda_c^+K^-\pi^+\pi^+$ and  $\Xi_{cc}^{++}\to\Xi_{c}^{+}\pi^{+}$ as the  golden discovery channels with $\Lambda_c^+$ reconstructed by $pK^-\pi^+$
and $\Xi_{c}^+ \to p K^- \pi^+$, respectively.
  %$\Xi_{cc}^+\to\Lambda_c^{+}\overline K^{*0}$ is the prior discovery channel for $\Xi_{cc}^+$.
}

\section{Motivation}

 Doubly or triply heavy flavor baryons with two or three heavy quarks ($b$ or $c$ quark) are predicted by the quark model, whose existence is also allowed by the QCD theory. However, the experimental search of these states is very slow. The first evidence was reported by the SELEX experiment for $\Xi_{cc}^+$ in 2002 \cite{Mattson:2002vu,Ocherashvili:2004hi}. However, it has never been  confirmed by later experiments   with larger data, such as FOCUS \cite{Ratti:2003ez}, BaBar \cite{Aubert:2006qw} and Belle \cite{Chistov:2006zj,Kato:2013ynr}. Utilizing as large as 0.65\,fb$^{-1}$ data, the LHCb collaboration even performed a thorough search in the discovery channel used by SELEX, $\Xi_{cc}^{+}\to \Lambda_{c}^{+}K^{-}\pi^{+}$, but did not find any significant signal \cite{Aaij:2013voa}. On the theoretical side, analysis on the production of doubly heavy baryons \cite{Zhang:2011hi,Chang:2005bf} indicates a large possibility of observing $\Xi_{cc}$ at LHC. Therefore searching for the doubly heavy baryons was proposed as an important physical goal by the LHCb collaboration. Although there had been a lot of research on the masses and decay constants of doubly heavy baryons, people knew little about doubly heavy baryon decays. To make the experimental searching more efficient, it became urgent and necessary to study the branching ratios of doubly heavy baryon decays.

As ground states of doubly heavy baryons, they can only decay weakly. There are a huge number of decay channels to study, since there are many hadronic states below their mass scale.  After a carful study of all possible decay channels of the doubly heavy baryons, we give suggestions of some golden channels with large branching ratios and all charged final sates for experimental search.\cite{Yu:2017zst} Following our suggestions, the LHCb experiment \cite{Aaij:2017ueg} did find the  $\Xi_{cc}^{++}$  state through one of our suggested decay channels $\Xi_{cc}^{++}\to \Lambda_c^+K^-\pi^+\pi^+$   with $\Lambda_c^+$ reconstructed by $pK^-\pi^+$.

 \section{Theoretical study of  doubly heavy baryons decays}
 %\subsection{Lifetimes}

 Comparing to the doubly heavy baryons with $b$ quark(s), the doubly charm baryons are easier to be observed because their production needs less energy. Under the flavor $SU(3)$ symmetry doubly charm baryons $\Xi_{cc}^{++}$, $\Xi_{cc}^{+}$ and $\Omega_{cc}^{+}$ form a spin-$\frac{1}{2}$ and a spin-$\frac{3}{2}$ triplets. The latter triplet will decay to the former one via electromagnetic or strong interactions. As ground states, the spin-$\frac{1}{2}$ triplet can only decay weakly through one of  the $c$ quarks.

 Lifetime is an important point in the choice of the candidates. On one hand the branching ratios are related directly to the lifetime of the mother particle, on the other hand particles with longer lifetimes will be easier to be identified in experiments. Different theoretical work gives quite different predictions for the lifetimes of doubly charm baryons \cite{Karliner:2014gca}, which makes it hard to judge by the absolute values of lifetimes. Luckily, because of the effect of destructive Pauli interference it is expected that $\tau(\Xi_{cc}^{++})\gg\tau(\Xi_{cc}^+)\sim \tau(\Omega_{cc})$. Their ratio is predicted \cite{Karliner:2014gca} to be
$\mathcal{R}_\tau\equiv {\tau_{\Xi_{cc}^+} / \tau_{\Xi_{cc}^{++}}} =0.25\sim0.37$ with less theoretical uncertainty.
Therefore we suggested $\Xi_{cc}^{++}$ as a prior candidate in experimental search.
This is contrary to the finding of  the SELLEX experiment, who declared the discovering of $\Xi_{cc}^{+}$ instead of $\Xi_{cc}^{++}$.

\subsection{Semileptonic decays}

\begin{figure}[htp]
\begin{center}
\begin{minipage}{0.33\linewidth}
\centerline{\includegraphics[scale=0.3]{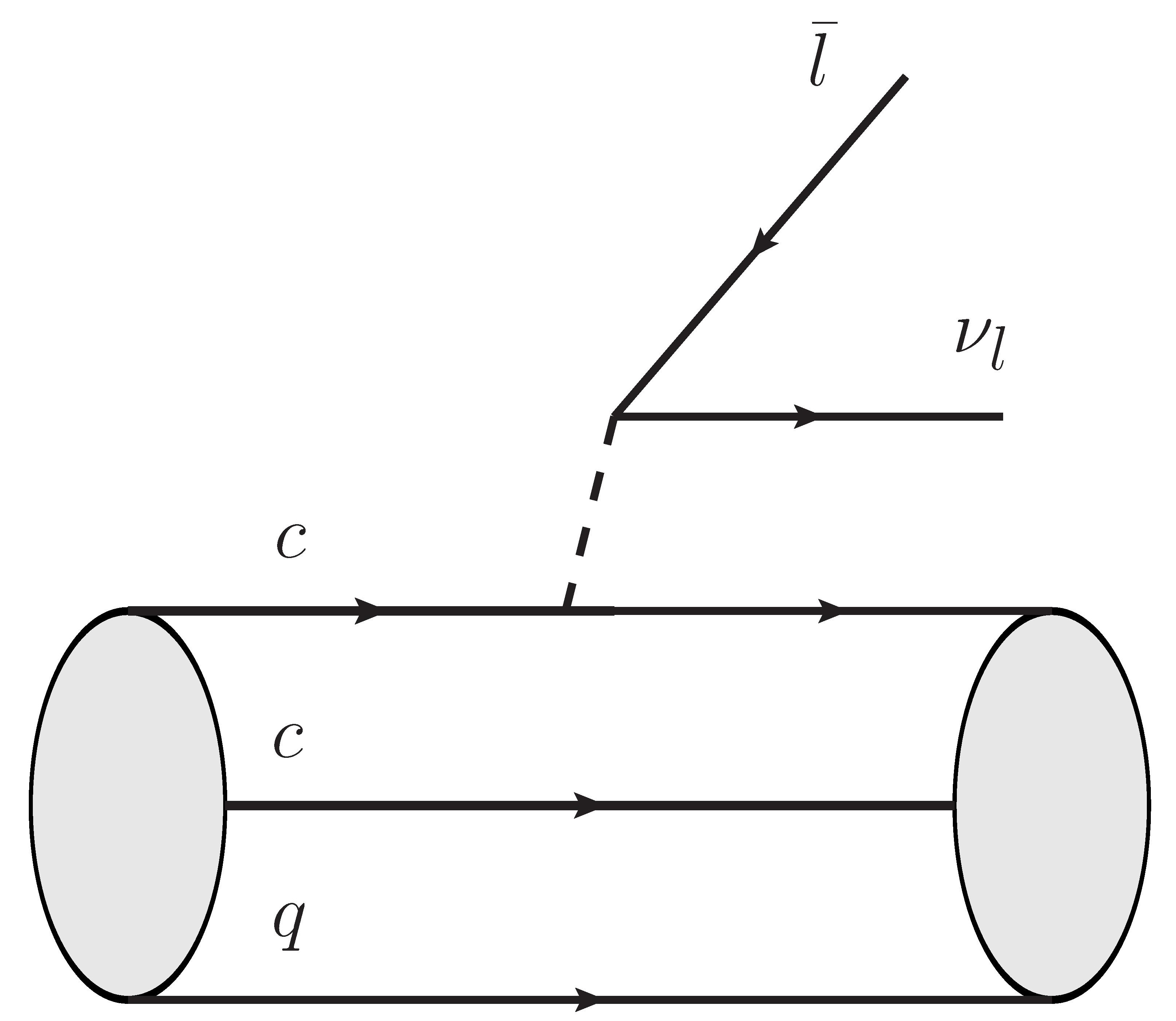}}
\caption{Semileptonic decay of a doubly charm baryon.}
\label{fig:semileptonic}
\end{minipage}
\hspace{2cm}
\begin{minipage}{0.33\linewidth}
\centerline{\includegraphics[scale=0.4]{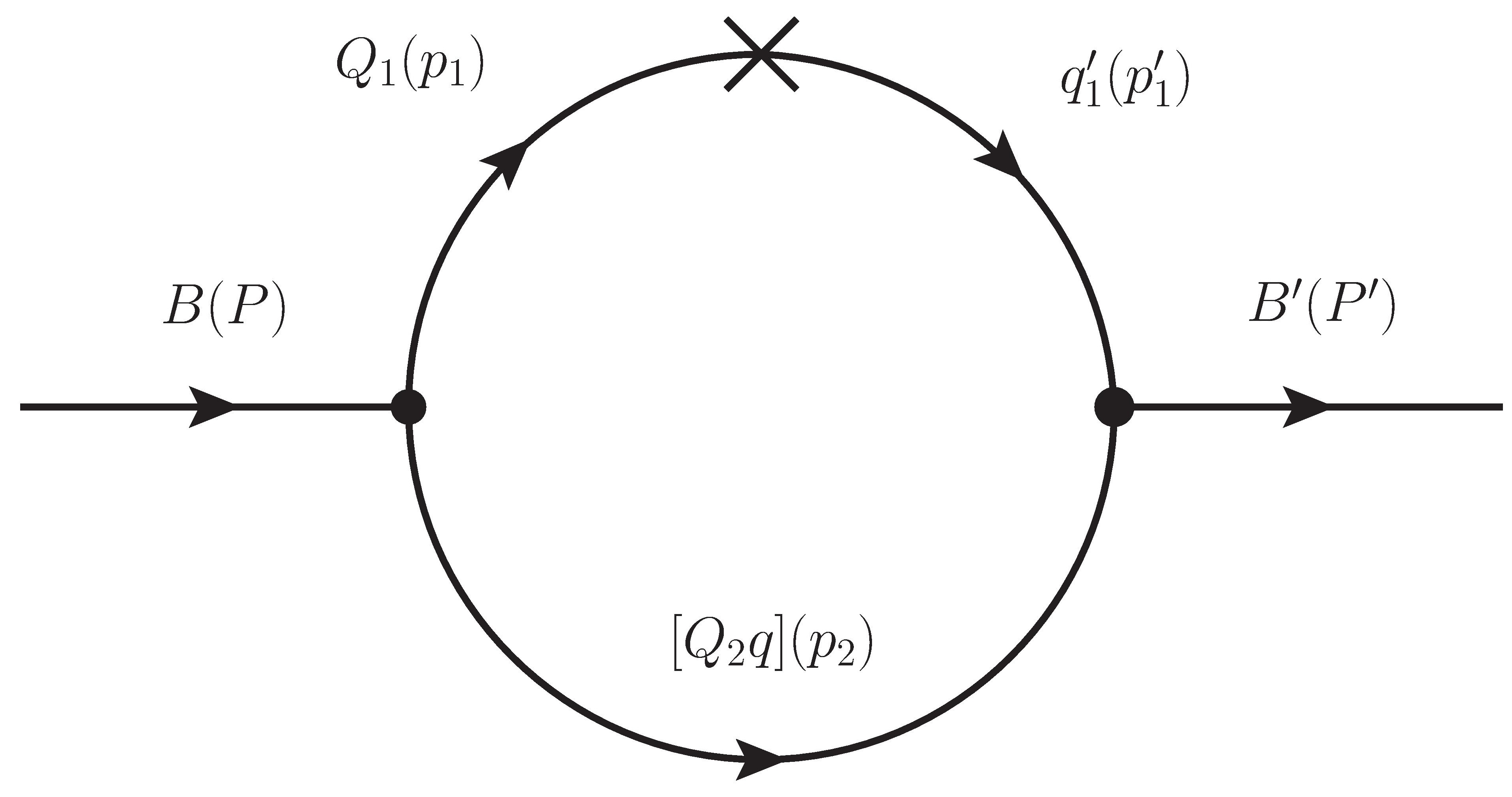}}
\caption{Baryon to baryon transition depicted by light front quark model in the diquark picture.}
\label{fig:LFQM}
\end{minipage}
\end{center}
\end{figure}

On the theoretical side the simplest case is semi-leptonic decays (shown in Fig. \ref{fig:semileptonic}), whose amplitudes can be factorized safely into a leptonic and a hadronic part. The hadronic part is a baryon to baryon weak transition matrix element, which can be parameterized as
{\small
\bea
\langle B^{\prime}(P^{\prime},S_{z}^{\prime})|(V-A)_{\mu}|B(P,S_{z})\rangle=&
\bar{u}(P^{\prime},S_{z}^{\prime})\left[\gamma_{\mu}f_{1}(q^{2})+i\sigma_{\mu\nu}\frac{q^{\nu}}{M}f_{2}(q^{2})+\frac{q_{\mu}}{M}f_{3}(q^{2})\right]u(P,S_{z})\nonumber \\
& - \bar{u}(P^{\prime},S_{z}^{\prime})\left[\gamma_{\mu}g_{1}(q^{2})+i\sigma_{\mu\nu}\frac{q^{\nu}}{M}g_{2}(q^{2})+\frac{q_{\mu}}{M}g_{3}(q^{2})\right]\gamma_{5}u(P,S_{z}).\label{eq:weakMatrix}
\eea
}
To get the branching ratios, the key task is the calculation of the form factors $f_{1,2,3}(q^2)$ and $g_{1,2,3}(q^2)$ defined in Eq.(\ref{eq:weakMatrix}). The first calculation is made       in the light front quark model (LFQM) \cite{Wang:2017mqp}. If one combines the spectators ($c$ and $q$ quarks) in Fig.~\ref{fig:semileptonic} as a diquark, a baryon to baryon transition is similar to a meson to meson one shown in Fig.\ref{fig:LFQM}. Since the diquark can be either a spin-$0$ or spin-$1$ state, a physical transition is a mixture of these two cases.
%\begin{equation}
%\langle B^{\prime}|(V-A)_{\mu}|B\rangle=c_{S}\langle q_{1}[Q_{2}q]_{S}|(V-A)_{\mu}|Q_{1}[Q_{2}q]_{S}\rangle+c_{A}\langle q_{1}[Q_{2}q]_{A}|(V-A)_{\mu}|Q_{1}[Q_{2}q]_{A}\rangle,\label{eq:csca}
%\end{equation}
%where the coefficients $c_{S,A}$ are determined by wave functions of the initial and final states.
With the form factors obtained in LFQM, the branching fractions of related semileptonic decays are all calculated.\cite{Wang:2017mqp} Here we only list the four branching ratios of $\Xi_{cc}^{++}$ semileptonic decays:
\bea
&{\cal BR}(\Xi_{cc}^{++}\to\Xi_{c}^{+}l^{+}\nu_{l})=5.25\times 10^{-2},\,
{\cal BR}(\Xi_{cc}^{++}\to\Xi_{c}^{\prime+}l^{+}\nu_{l})=5.84\times 10^{-2},\nonumber\\
&{\cal BR}(\Xi_{cc}^{++}\to\Lambda_{c}^{+}l^{+}\nu_{l})=4.81\times 10^{-3} ,\,
{\cal BR}(\Xi_{cc}^{++}\to\Sigma_{c}^{+}l^{+}\nu_{l})=4.38\times 10^{-3},
\label{eq:Brofsemi}
\eea
where the two decays in the first line of Eq. (\ref{eq:Brofsemi}) are induced by $c \to s$ transition and those in the second line by $c \to d$ transition. One can see that branching ratios of the two Cabbibo favored decays are at the order of $10^{-2}$ which is about $10$ times larger than the Cabbibo suppressed ones. For the sake of particle reconstruction in experiments one needs to take the secondary decays into consideration, which decreases the branching ratios by a factor of $10^{-2}$.
As a result, the branching ratios of semi-leptonic decays used in experimental research are expected to be of the order $10^{-4}$. Comparing to non-leptonic decays, whose branching ratios\cite{Li:2017ndo} are of the order $10^{-3} \sim 10^{-4}$, the semi-leptonic decays are not competitive. what's more, the largest disadvantage of semi-leptonic decays is the phenomena of missing energy in experiments caused by the neutrinos.

\subsection{Non-leptonic decays}

\begin{figure}[htp]
\begin{center}
\hspace{2cm}
\begin{minipage}{0.33\linewidth}
\centerline{\includegraphics[scale=0.42]{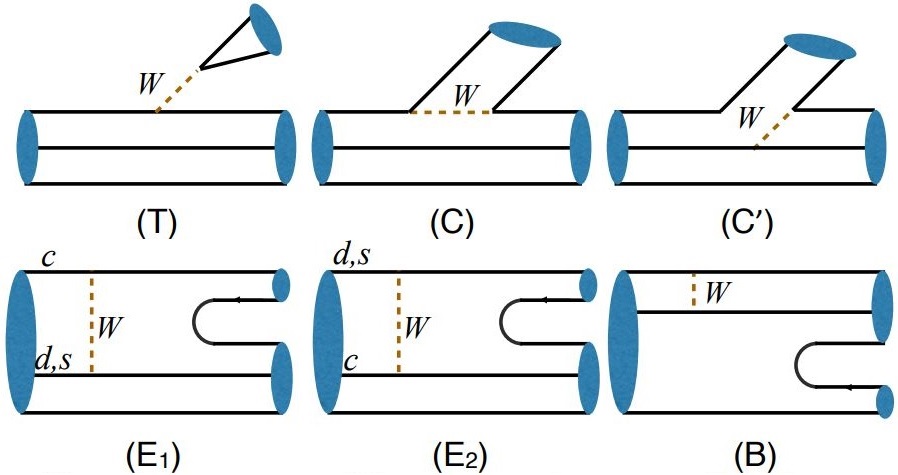}}
\caption{Topologies of two body nonleptonic decays of doubly charm baryons.}
\label{fig:tbdecays}
\end{minipage}
\hspace{3cm}
\begin{minipage}{0.33\linewidth}
\centerline{\includegraphics[scale=0.17]{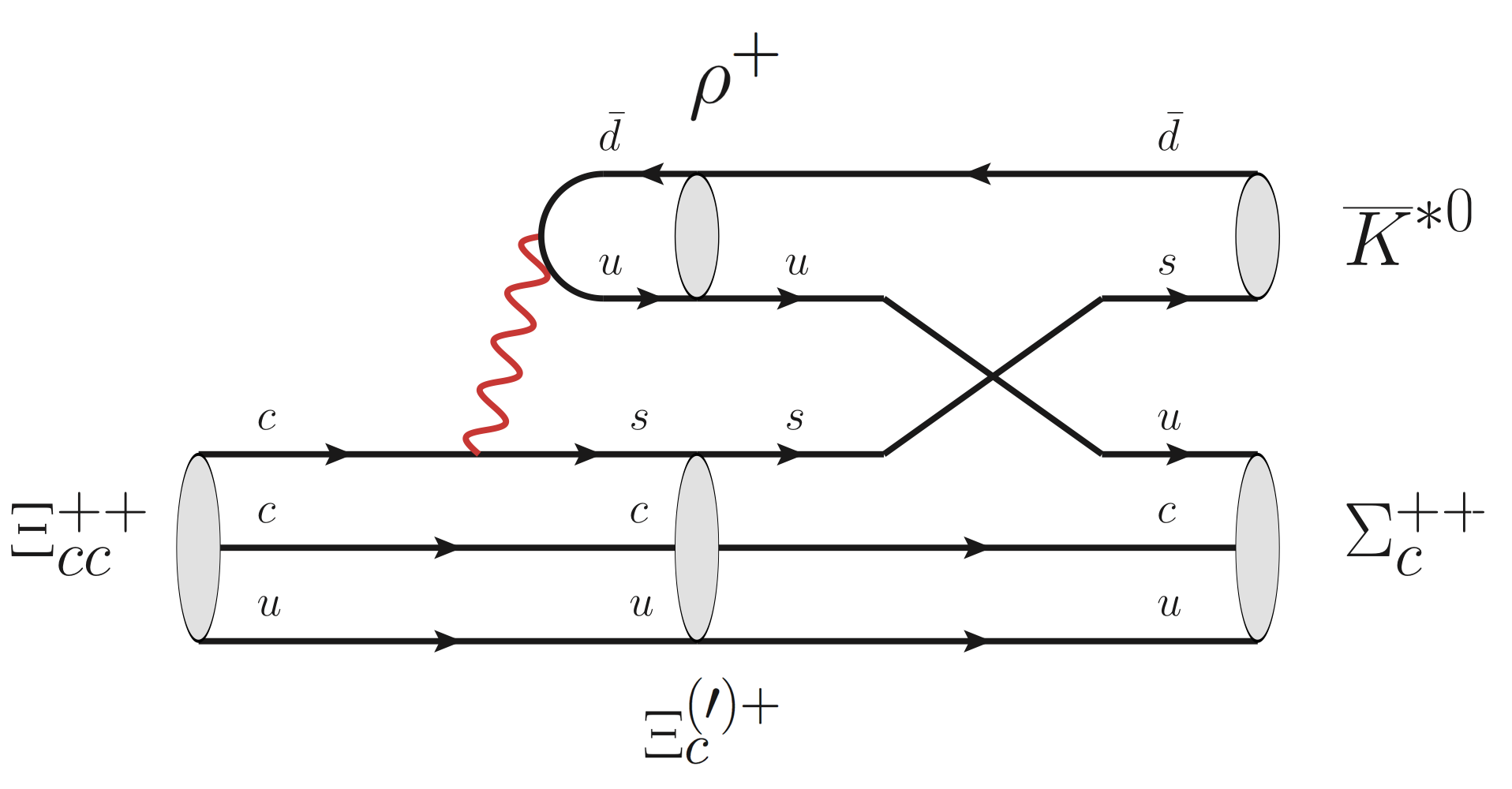}}
\centerline{\includegraphics[scale=0.17]{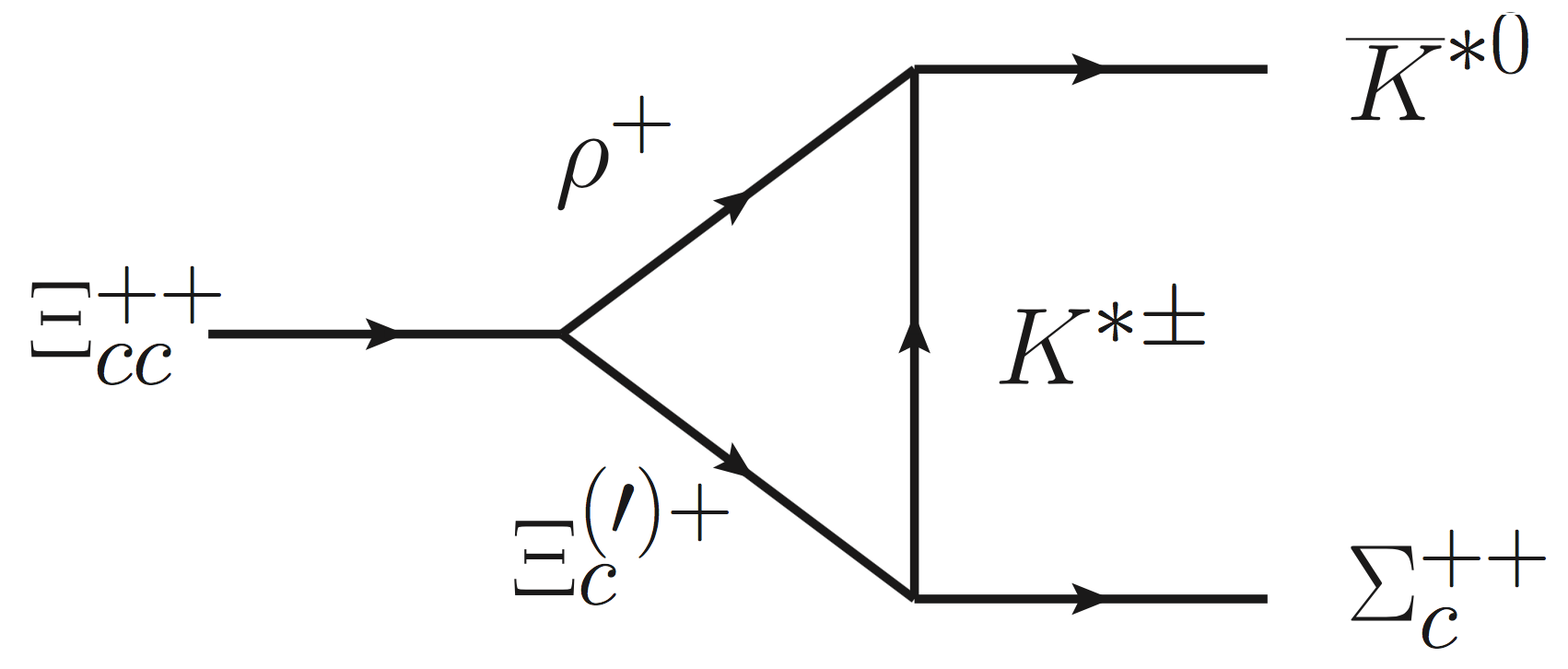}}
\caption{\label{fig:triangle} The rescattering diagram of $\Xi_{cc}^{++}\to \Xi_{c}^{(\prime)+}\rho^{+}\to\Sigma_c^{++}\overline K^{*0}$. Shown at quark and hadron level.}
\end{minipage}
\end{center}
\end{figure}

 We first consider the simplest non-leptonic decays: a doubly charm baryon decays into a single charm baryon and a meson. Theoretically non-leptonic charm decays are difficult to deal with. The scale of charm is very special, which is much higher than $\Lambda_{\rm{QCD}}$ but not high enough for a good  heavy quark mass expansion. The non-perturbative QCD plays   a major role, which restricts the application of most factorization theories. The factorization assisted topological diagrammatic approach, which works well in charm meson decays to predict the $\Delta A_{\mbox{cp}}$ successfully \cite{Li:2012cfa}, also does not work here, since there are not enough experimental data. Therefore the doubly charm baryon decays need to be considered from the beginning. The leading order topological diagrams contributions in decays of doubly charm baryon to a single charm baryon and a meson can be classified into the diagrams in Fig. \ref{fig:tbdecays}.\cite{Yu:2017zst}

 From the experience of D meson decays,\cite{Li:2012cfa} we notice that the emission diagram (denoted by T) is dominated by the short distance contribution, which can be factorized as production of a meson decay constant and a weak transition of baryon to baryon. With the form factors evaluated in LFQM,\cite{Wang:2017mqp} this  contribution is easy to    calculate.
 Considering the detection efficiency and comparing among the theoretical results, we find that among short distance contribution denominated decays $\Xi_{cc}^{++}\to\Xi_{c}^{+}\pi^{+}$ is the best channel for detection with $\Xi_{c}^+$ reconstructed by the final state $p K^- \pi^+$. The secondary decay $\Xi_{c}^+ \to p K^-\pi^{+}$ is estimated to have a branching ratio of about two percent.\cite{Yu:2017zst}

Similar to the D meson decays,  long distance contribution is    dominating in other diagrams. This part of contribution has never been calculated in the literature and we evaluate it firstly with the rescattering mechanism of the final-state-interaction. The rescattering mechanism is induced by quark exchanges, and it can be calculated as triangle diagrams by using the effective Lagrangian with couplings at hadron level. The depiction of this mechanism is shown in Fig. \ref{fig:triangle} with the $t$-channel rescattering diagram of $\Xi_{cc}^{++}\to \Xi_{c}^{(\prime)+}\rho^{+}\to\Sigma_c^{++}\overline K^{*0}$ as an example. The calculated  branching ratios are expected to have large errors because of the uncertain hadronic parameters. In order to reduce ambiguity, we use their ratios instead of branching fractions to draw conclusions. Most of the theoretical uncertainties due to the hadronic parameters canceled in the ratios. Our sample results of ratios of the long distance dominated decays are listed in Table \ref{tab:ratios}.

%------------------------------------------------------------
\begin{table}[htp]
\begin{center}
\caption{\label{tab:ratios}Branching fractions of $\Xi_{cc}^{++}$ and $\Xi_{cc}^+$ decays with the long-distance contributions, relative to that of $\Xi_{cc}^{++}\to \Sigma_{c}^{++}(2455)\overline K^{*0}$.}

\begin{tabular*}{84mm}{@{\extracolsep{\fill}}ccc}
\hline
Baryons & \quad\quad\quad\quad Modes \quad\quad\quad\quad & $\mathcal{B}_{\rm LD}$
\\
\hline
$\Xi_{cc}^{++}(ccu)$ &$\Sigma_{c}^{++}(2455)\overline K^{*0}$ &  defined as 1
\\
&$pD^{*+}$  & $0.04$
\\
&$pD^+$  & $0.0008$
\\
\hline
%Baryons & Modes & $\mathcal{B}_{\rm LD}(\times\tau_{\Xi_{cc}^{+}}/100\text{fs})$ &\\\hline
$\Xi_{cc}^+(ccd)$ &$\Lambda_c^+\overline K^{*0}$ &  $(\mathcal{R}_{\tau}/0.3) \times 0.22$
\\
&$\Sigma_c^{++}(2455)K^-$ &  $(\mathcal{R}_{\tau}/0.3) \times 0.01$
\\
&$\Xi_c^+\rho^0$ &  $(\mathcal{R}_{\tau}/0.3) \times 0.04$
\\
& $\Lambda D^+$ & $(\mathcal{R}_{\tau}/0.3) \times 0.004$
\\
&$pD^0$ &  $(\mathcal{R}_{\tau}/0.3) \times 0.001$
\\
\hline
\end{tabular*}
\end{center}
\end{table}
%------------------------------------------------------------

We find that $\Xi_{cc}^{++}\to\Sigma_c^{++}\overline K^{*0}$ decay has the largest branching fraction at the order of several percent. The secondary decays can be $\Sigma_c^{++}\to \Lambda_c^+ \pi^+$ and $\overline K^{*0}\to K^- \pi^+$.
The total 4-body decay is very difficult to calculate precisely, but our estimation shows that ${\cal BR}(\Xi_{cc}^{++}\to \Lambda_c^+ K^- \pi^+ \pi^+)$ can even reach ${\cal O}(10\%)$ because of the rich resonant contributions. In detection, $\Lambda_c$ can be reconstructed by $pK^-\pi^+$.
%One can see in Table \ref{tab:ratios} that $\Xi_{cc}^+\to\Lambda_c^{+}\overline K^{*0}$ can be chosen as the discovery decay of $\Xi_{cc}^+$.

\section{Conclusion}

In the purpose of finding out the golden   decay channels for experimental search, we systematically analyzed the properties and decays of doubly heavy baryons. Utilizing the form factors of doubly heavy baryon to single heavy baryon weak transition, the branching fractions of the semi-leptonic decays and the two body non-leptonic decays of doubly charm baryons are calculated. We find that $\Xi_{cc}^{++}\to\Lambda_c^+ K^- \pi^+ \pi^+$ with $\Lambda_c \to pK^-\pi^+$ has the first priority. In 2017, the LHCb collaboration declares the discovery of $\Xi_{cc}^{++}$ via this decay following our suggestions.\cite{Aaij:2017ueg} We also suggest the $\Xi_{cc}^{++}\to\Xi_{c}^{+}\pi^{+}$ decay with $\Xi_{c}^+ \to p K^-\pi^{+}$  as     a good candidate  for experimental search. As for  the   search of $\Xi_{cc}^+$ state, our calculation shows that $\Xi_{cc}^+\to\Lambda_c^{+}\overline K^{*0}(\to K^-\pi^+)$ is a prior candidate.

\section*{Acknowledgments}

We thank F.S.~Yu, H.Y.~Jiang,   W.~Wang and Z.X.~Zhao for excellent collaboration. This work was partly supported by National Natural Science Foundation of China (Grant
Nos. 11521505, 11621131001, 11447009, 11505098, and 11765012), and partly supported by the plan of \emph{Young Creative Talents} under \emph{the Talent of the Prairie} project of Inner Mongolia.


\section*{References}

\begin{thebibliography}{99}

%\cite{Mattson:2002vu}
\bibitem{Mattson:2002vu}
  M.~Mattson {\it et al.} [SELEX Collaboration],
  %``First observation of the doubly charmed baryon Xi+(cc),''
  Phys.\ Rev.\ Lett.\  {\bf 89}, 112001 (2002).

%\cite{Ocherashvili:2004hi}
\bibitem{Ocherashvili:2004hi}
  A.~Ocherashvili {\it et al.} [SELEX Collaboration],
  %``Confirmation of the double charm baryon Xi+(cc)(3520) via its decay to p D+ K-,''
  Phys.\ Lett.\ B {\bf 628}, 18 (2005).
  %%CITATION = doi:10.1016/j.physletb.2005.09.043;%%
  %239 citations counted in INSPIRE as of 24 Feb 2018


%\cite{Ratti:2003ez}
\bibitem{Ratti:2003ez}
  S.~P.~Ratti,
  %``New results on c-baryons and a search for cc-baryons in FOCUS,''
  Nucl.\ Phys.\ Proc.\ Suppl.\  {\bf 115}, 33 (2003).
  %%CITATION = doi:10.1016/S0920-5632(02)01948-5;%%
  %57 citations counted in INSPIRE as of 24 Feb 2018


%\cite{Aubert:2006qw}
\bibitem{Aubert:2006qw}
  B.~Aubert {\it et al.} [BaBar Collaboration],
  %``Search for doubly charmed baryons Xi(cc)+ and Xi(cc)++ in BABAR,''
  Phys.\ Rev.\ D {\bf 74}, 011103 (2006).
  %%CITATION = doi:10.1103/PhysRevD.74.011103;%%
  %128 citations counted in INSPIRE as of 24 Feb 2018


%\cite{Chistov:2006zj}
\bibitem{Chistov:2006zj}
  R.~Chistov {\it et al.} [Belle Collaboration],
  %``Observation of new states decaying into Lambda(c)+ K- pi+ and Lambda(c)+ K0(S) pi-,''
  Phys.\ Rev.\ Lett.\  {\bf 97}, 162001 (2006).
  %%CITATION = doi:10.1103/PhysRevLett.97.162001;%%
  %197 citations counted in INSPIRE as of 24 Feb 2018


%\cite{Kato:2013ynr}
\bibitem{Kato:2013ynr}
  Y.~Kato {\it et al.} [Belle Collaboration],
  %``Search for doubly charmed baryons and study of charmed strange baryons at Belle,''
  Phys.\ Rev.\ D {\bf 89}, no. 5, 052003 (2014).
  %%CITATION = doi:10.1103/PhysRevD.89.052003;%%
  %55 citations counted in INSPIRE as of 24 Feb 2018


%\cite{Aaij:2013voa}
\bibitem{Aaij:2013voa}
  R.~Aaij {\it et al.} [LHCb Collaboration],
  %``Search for the doubly charmed baryon $\Xi_{cc}^+$,''
  JHEP {\bf 1312}, 090 (2013).
  %%CITATION = doi:10.1007/JHEP12(2013)090;%%
  %58 citations counted in INSPIRE as of 24 Feb 2018

%\cite{Zhang:2011hi}
\bibitem{Zhang:2011hi}
  J.~W.~Zhang, X.~G.~Wu, T.~Zhong, Y.~Yu and Z.~Y.~Fang,
  %``Hadronic Production of the Doubly Heavy Baryon $\Xi_{bc}$ at LHC,''
  Phys.\ Rev.\ D {\bf 83}, 034026 (2011).
  %%CITATION = doi:10.1103/PhysRevD.83.034026;%%
  %15 citations counted in INSPIRE as of 24 Feb 2018


%\cite{Chang:2005bf}
\bibitem{Chang:2005bf}
  C.~H.~Chang, C.~F.~Qiao, J.~X.~Wang and X.~G.~Wu,
  %``The Color-octet contributions to P-wave $B_c$ meson hadroproduction,''
  Phys.\ Rev.\ D {\bf 71}, 074012 (2005).
  %%CITATION = doi:10.1103/PhysRevD.71.074012;%%
  %35 citations counted in INSPIRE as of 24 Feb 2018
%\cite{Aaij:2017ueg}

%\cite{Yu:2017zst}
\bibitem{Yu:2017zst}
  F.~S.~Yu, H.~Y.~Jiang, R.~H.~Li, C.~D.~Lu, W.~Wang and Z.~X.~Zhao,
  %``Discovery Potentials of Doubly Charmed Baryons,''
Chinese Phys. C42 (2018) 051001
\bibitem{Aaij:2017ueg}
  R.~Aaij {\it et al.} [LHCb Collaboration],
  %``Observation of the doubly charmed baryon $\Xi_{cc}^{++}$,''
  Phys.\ Rev.\ Lett.\  {\bf 119}, no. 11, 112001 (2017).

%\cite{Karliner:2014gca}
\bibitem{Karliner:2014gca}
  M.~Karliner and J.~L.~Rosner,
  %``Baryons with two heavy quarks: Masses, production, decays, and detection,''
  Phys.\ Rev.\ D {\bf 90}, no. 9, 094007 (2014) and references therein
  %%CITATION = doi:10.1103/PhysRevD.90.094007;%%
  %56 citations counted in INSPIRE as of 24 Feb 2018

 %\cite{Guberina:1999mx}
%\bibitem{Guberina:1999mx}
  %B.~Guberina, B.~Melic and H.~Stefancic,
  %``Inclusive decays and lifetimes of doubly charmed baryons,''
  %Eur.\ Phys.\ J.\ C {\bf 9}, 213 (1999)
  %[Eur.\ Phys.\ J.\ C {\bf 13}, 551 (2000)].
  %%CITATION = doi:10.1007/s100529900039, 10.1007/s100520050525;%%
  %48 citations counted in INSPIRE as of 24 Feb 2018

%\cite{Wang:2017mqp}
\bibitem{Wang:2017mqp}
  W.~Wang, F.~S.~Yu and Z.~X.~Zhao,
  %``Weak decays of doubly heavy baryons: the $1/2\rightarrow 1/2$ case,''
  Eur.\ Phys.\ J.\ C {\bf 77}, 781 (2017).
  %%CITATION = doi:10.1140/epjc/s10052-017-5360-1;%%
  %22 citations counted in INSPIRE as of 05 May 2018

%\cite{Li:2017ndo}
\bibitem{Li:2017ndo}
  R.~H.~Li, C.~D.~Lu, W.~Wang, F.~S.~Yu and Z.~T.~Zou,
  %``Doubly-heavy baryon weak decays: $\Xi_{bc}^{0}\to pK^{-}$ and $\Xi_{cc}^{+}\to \Sigma_{c}^{++}(2520)K^{-}$,''
  Phys.\ Lett.\ B {\bf 767}, 232 (2017).

%\cite{Li:2012cfa}
\bibitem{Li:2012cfa}
  H.~n.~Li, C.~D.~Lu and F.~S.~Yu,
  %``Branching ratios and direct CP asymmetries in $D\to PP$ decays,''
  Phys.\ Rev.\ D {\bf 86}, 036012 (2012).


\end{thebibliography}
\end{document}